# Aerothermodynamic Analysis of Faceted Aeroshell at Hypersonic Speed

Pietro Innocenzi [1], Michela Gramola [1], Tom B. Fisher [2],
Mark K. Quinn [2], Paul J. K. Bruce [1], Salvador Navarro-Martinez [1]

## Abstract

This study explores the aerothermal behaviour of a rigid mechanically deployable aeroshell developed at Imperial College London for high-payload atmospheric entry missions. The multiphysics CFD software STAR-CCM+ is used to perform a Conjugate Heat Transfer analysis on the aeroshell's faceted geometry. Results are presented for four different geometry models tested in air at Mach 5 with angles of attack α = 0°, 5° and 10°. The predicted surface heat transfer reveals areas of elevated heat loads at the ribs between facets and at the aeroshell shoulder, due to local boundary layer thinning. The increase in heat transfer at the ribs depends on the sharpness of the rib: more rounded shapes result in lower heat fluxes. Comparison with high-speed wind tunnel tests shows good agreement with experimental data. Stanton number and temperature profiles agree within 8% and 2% respectively. The discrepancies between experiments and simulations are largest at the sharp ribs of the aeroshell. The sources of error can be associated to three dimensional effects neglected in the heat flux derivations from temperature measurements as well as experimental uncertaintes.

## Nomenclature

$C_H$ – Stanton number
$d$ – Article diameter
$H_0$ – Total enthalpy
$h_w$ – Wall enthalpy
$M$ – Mach number
$P_0$ – Total pressure
$q_w$ – Wall convective heat transfer
$Re_d$ – Reynolds number based on diameter
$Re_c$ – First cell Reynolds number

$T_0$ – Total temperature
$T_{CHT}$ – Temperature obtained with CHT
$T_{exp}$ – Temperature obtained with experiment
$T_W$ – Wall temperature
$u_\infty$ – Freestream velocity
$\alpha$ – Angle of attack
$\Delta t$ – Time-step size
$\varepsilon_T$ – Temperature error
$\rho_\infty$ – Freestream density

## 1. Introduction

To enable future exploration of Mars and the return of samples from other planets to specific locations on Earth, next generation space missions will require higher payload mass than current and past missions. To achieve low ballistic coefficients, an increase in aeroshell frontal area is needed. However, current launcher fairings cannot stow aeroshells of diameter larger than 4.6 m [1].
Deployable aeroshells have the potential to deliver an equivalent payload with a stowed diameter up to four times smaller than that of a rigid aeroshell [2] and are an attractive technology for future entry, descent, and landing missions. A robust and versatile option for deployable aerothermal shields is the mechanically deployable concept [3-5], consisting of a set of rigid radial ribs used to deploy the shield, see Fig. 1 [5].

---

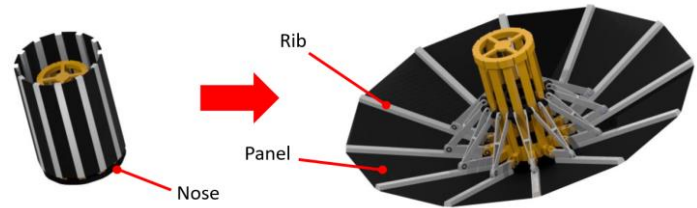

**Fig. 1:** Mechanically deployable aeroshell [5]

Research carried out at Imperial College London focuses on the design and development of a novel rigid mechanically deployable aeroshell: the Hypersonic foldable Aeroshell for THermal protection using ORigami (HATHOR) [5]. HATHOR consists of rigid panels with a thermal protection system layer connected between 8 retractable ribs allowing for a deployed diameter of 2.65 m, see Fig. 2 [5]. The rigid panels can be folded according to an optimal pattern following origami principles, allowing compact stowage and high precision deployment.

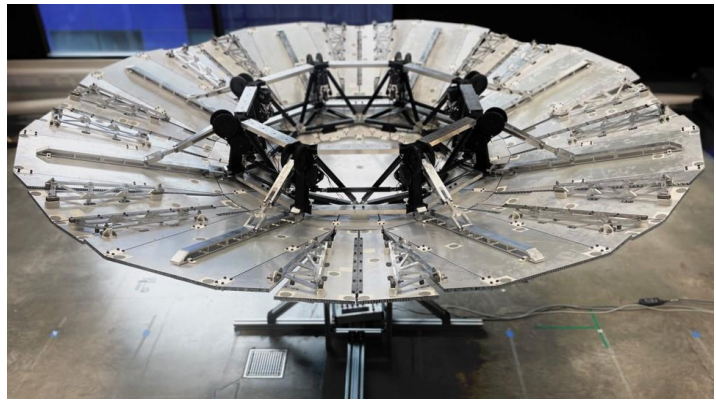

**Fig. 2:** HATHOR engineering demonstrator model [5]

Studies of hypersonic aerothermal analyses of faceted aeroshells are scarce, although some do exist [6-8]. Existing CFD studies [6, 7] compute the heat fluxes solving the energy equation only for the fluid, imposing a cold wall or radiative cooling boundary condition at the surface. However, such boundary conditions are non-physical, as the surface temperature is expected to increase in time. In addition, these simulations do not consider the thermal conduction through the solid, which can be considerable near corners and edges. A Conjugate Heat Transfer (CHT) analysis can be performed to model this, coupling the fluid energy equation to the solid heat conduction equation. The objective of this study is to carry out a CHT analysis to analyse the aerothermal behaviour of HATHOR's faceted geometry in the hypersonic laminar regime. The commercial CFD solver STAR-CCM+ can simulate hypersonic flows with comparable accuracy to research-oriented flow solvers for two-dimensional and axisymmetric cases [9]. This study also aims at validating the solver capability in predicting the heating of three-dimensional faceted geometries exposed to hypersonic laminar fluxes.

## 2. Methodology

### 2.1. Aeroshell geometry

To understand the aerothermal behaviour of faceted heat shields, four different models are analysed, see Fig. 3a. The *Sharp* model resembles the frontal shape of HATHOR, with eight facets connected between perfectly sharp ribs. The *Shoulder* model has an identical frontal face, but is characterised by a partially open backshell, to study the influence of low heatshield thickness on the results. The *Smooth* geometry is similar to the *Sharp* one, but is characterised by smooth ribs, to analyse the influence of ribs sharpness on the flow. Finally, the *Sphere-cone* geometry is used as a reference and represents the frontal geometry of a conventional entry vehicle. A campaign of wind tunnel tests has been performed at the University of Manchester's High SuperSonic Tunnel to obtain heat flux results. The four different models with diameter d = 30 mm were manufactured from MACOR®, a machinable glass ceramic with very low thermal diffusivity and high emissivity. The models are shown in Fig. 3b.

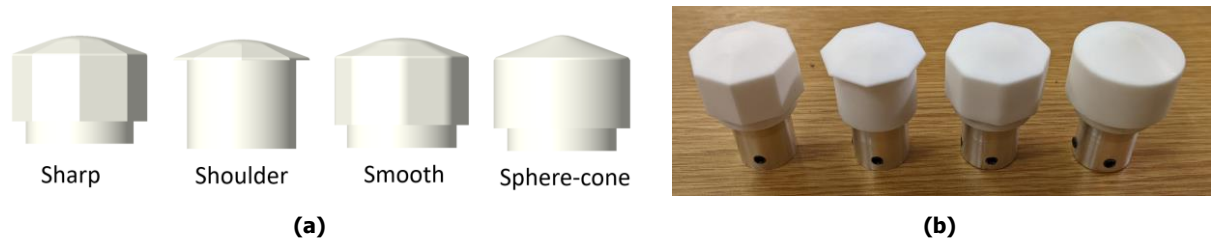

(a)                                           (b)

**Fig. 3:** Aeroshell geometries used in this study: (a) CAD models, (b) manufactured models

## 2.2. Flow solver and models

In this work the conservation equations of mass, momentum, and energy for the fluid are solved together with the heat conduction equation for the solid in a fully coupled manner. The flux discretization method is the MUSCL central differencing scheme [10] with 3$^{rd}$ order accuracy in space. The method used to evaluate the inviscid fluxes is the AUSM+ flux-vector splitting scheme [11], and the grid sequencing initialization feature is used to obtain fast residuals convergence.

The unsteady equations employ a 2$^{nd}$ order discretization scheme in time. A steady simulation only for the fluid is first performed. Once the fluid has reached steady state, both fluid and solid equations are solved simultaneously in a transient simulation with time-step size $\Delta t = 0.01$ s. Shorter time-steps did not lead to different results. This methodology is justified by the fact that the vehicle's wake is the only area characterized by unsteady flow, and the frontal faceted geometry – which is the focus of this study – is not influenced by its time-accurate computation. In addition, the influence induced by the solid transient on the flowfield at the wall is limited and can be neglected. This is demonstrated by the fact that shorter time-steps did not lead to different results in heat flux.

The working fluid is air, which is modelled as a real gas. Air dissociation, air ionization and non-equilibrium thermodynamic effects are not modelled due to the relatively low Mach number. The compressibility factor, specific heat, enthalpy, viscosity, and thermal conductivity vary as a function of temperature and pressure as specified by the STAR-CCM+ Equilibrium Air model.

## 2.3. Boundary conditions and mesh

The models are included in a cylindrical domain of diameter 10d and length 20d. A sting is attached to the back of each model to simulate wind tunnel conditions. To improve the numerical stability of the code, and to account for the impossibility of manufacturing perfectly sharp edges, the ribs of the aeroshells named *Sharp* and *Shoulder* have a fillet of radius 0.7 mm. The inlet flow conditions of M = 5, $T_0$ = 800 K and $P_0$ = 830 kPa are selected to match wind tunnel conditions, leading to $Re_d = 1.5 \times 10^5$. Due to the relatively low Reynolds number, only laminar simulations are performed. During the steady CFD simulation, the solid surface boundary condition is a non-slip isothermal wall condition with uniform $T_w$ = 300 K. Once the transient CHT simulation is started, this condition is removed and the surface temperature increases approaching the adiabatic wall temperature. The four different aeroshell geometries are simulated at angles of attack α = 0°, 5°, 10°. In each case, the mesh is generated within STAR-CCM+ and consists of prism layer cells near viscous boundaries surrounded by polyhedral mesh elsewhere. The entire aeroshell geometry is meshed and simulated. Simulations with half aeroshell geometry were also performed but unphysical heat flux profiles raised at the symmetry planes. A mesh convergence study was first carried out leading to a final converged mesh of around 20 million cells for each test case. To obtain accurate results, particular care is given to both bow shock and boundary layer refinement. An adaptive mesh refinement based on normalized pressure jumps [9] is applied to refine the mesh across the bow shock and supersonic expansions. The first prism mesh layer on the front of the model is characterized by a first cell Reynolds number $Re_c < 12$ throughout, which proved to be enough to allow mesh independence on the heat flux results [12]. A section of the final mesh is shown in Fig. 4.

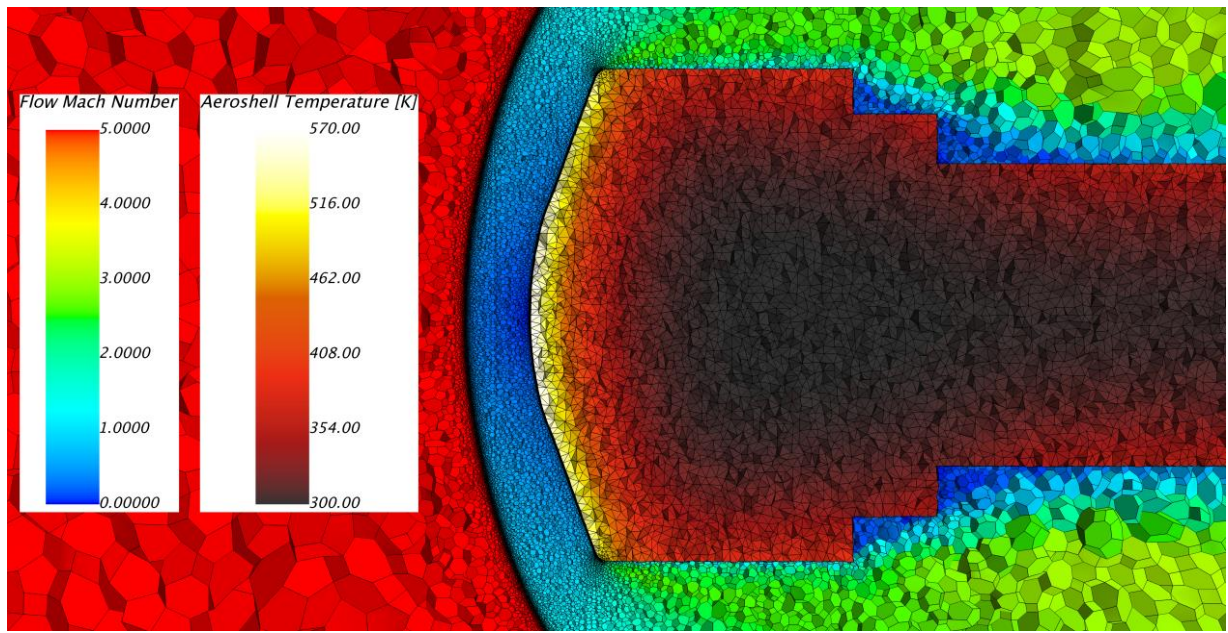

**Fig. 4:** Cross section along the centreplane of the 3D mesh (*Smooth* geometry)

## 2.4. Validation

Experiments were conducted at the University of Manchester's High SuperSonic Tunnel (HSST), a long duration blow-down facility. The tunnel allows for run times of up to 10 s, and the run time for the current tests was ~7 s. The experimental set-up is shown in Fig. 5a. The temperature histories are collected using infrared thermography (IRT). After camera calibration, each pixel's radiant intensity corresponds to a certain value of temperature, see Fig. 5b. The experimental heat fluxes are calculated from the temperatures solving the 1D heat equation with a direct numerical technique [13]. The experimentally measured wall temperature is applied as a Dirichlet boundary condition, and the problem is discretized with a finite volume method. A major assumption of the heat flux derivation is that of a semi-infinite body in the axial direction.

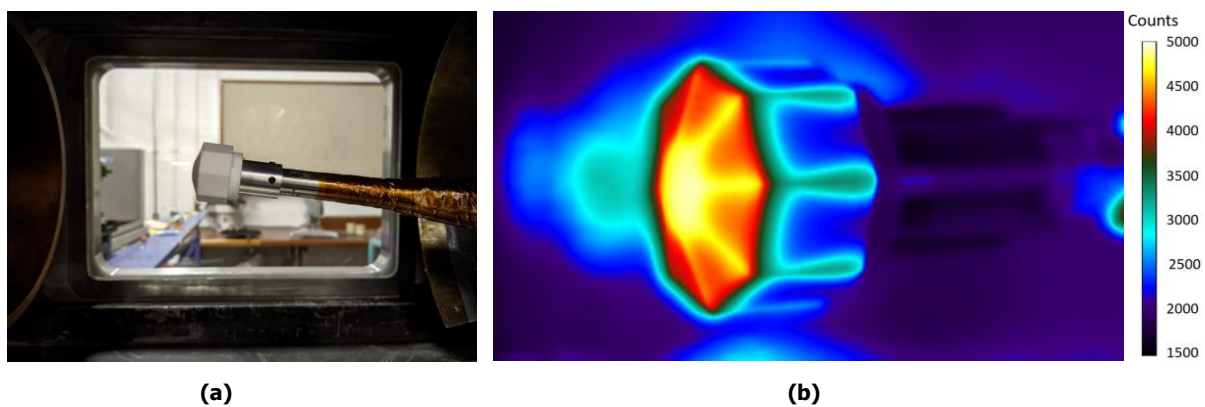

(a)  (b)

**Fig. 5:** (a) Experimental set-up in the HSST working section, (b) IRT radiant intensity output

## 3. Results and Discussion

### 3.1. Baseline Flow

Figure 6a shows the density contour for the *Smooth* model at α = 0°. The main features of a hypersonic flow around a blunt body are visible: the bow shock in front of the capsule, the supersonic expansions at the shoulder as well as a recirculation region near the wake. The absence of recompression shocks downstream the wake is attributed to the presence of the sting. Bow shock shape and shock stand-off distance show qualitative agreement with the experimental schlieren given in Fig. 6b.

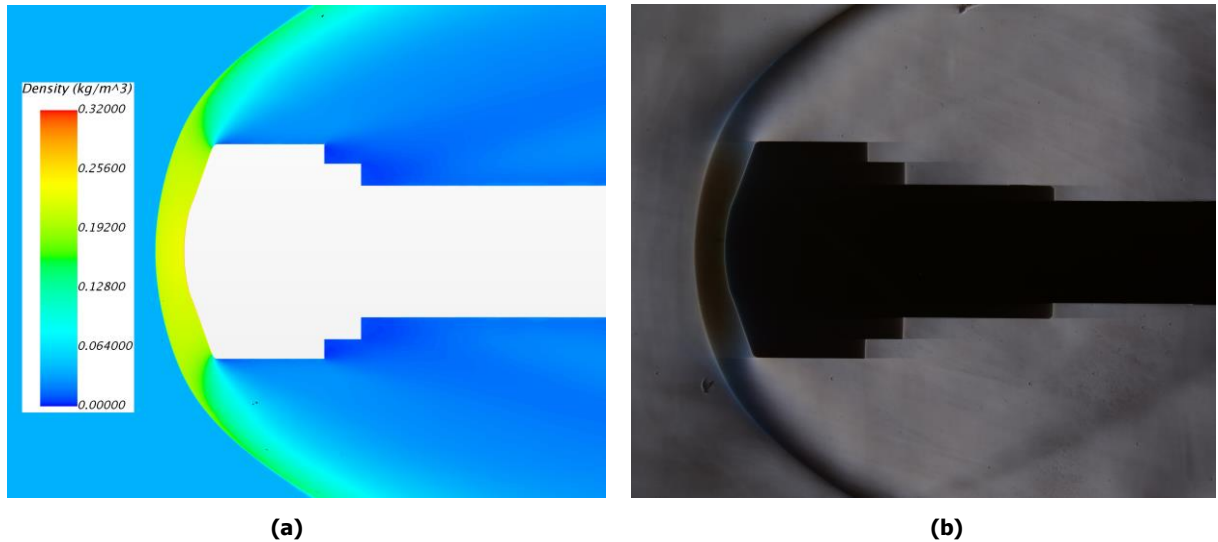

**(a)**        **(b)**

**Fig. 6:** Comparison of baseline flow for *Smooth* model at α = 0°: (a) cross-sectional density contour, (b) experimental schlieren

### 3.2. Effects of geometry and angle of attack

Heat flux results over the aeroshell surface are presented as modified Stanton number, defined as:

$$C_H = \frac{q_w}{\rho_\infty u_\infty (H_0 - h_w)} \qquad (1)$$

where $q_w$ is the convective heat transfer at the wall, $\rho_\infty$ and $u_\infty$ are freestream density and velocity, $H_0$ is total enthalpy and $h_w$ is static enthalpy at the wall.

Modified Stanton number contours at t = 5s and α = 0°, 5°, 10° are shown in Fig. 7-9 respectively, for both experimental tests and numerical simulations. Stanton number contours do not change significantly in time but are strongly affected by the geometry of the model. All the faceted geometries exhibit higher heating at the ribs than at the panels. As the air flows from the nose to the ribs, it is forced by the geometry to expand towards the panels leading to a decrease in boundary layer thickness at the rib, causing higher convective heating from the fluid to the body. The models with sharp ribs experience a 40% increase in Stanton number at the ribs compared to the facets. The smooth ribs exhibit only a 20% increase, indicating that the strength of this effect correlates with the level of sharpness of the rib, with smoother ribs leading to decreased boundary layer thinning, and a smaller increase in local heat flux. Such values compare reasonably with a previous laminar CFD study of NASA's ADEPT faceted aeroshell at high Mach numbers (between 10 and 30) [6], whose sharp ribs exhibited a 33% increase in convective heat transfer with respect to the facets. The cases α = 5° and 10° show asymmetric contours with increased heat flux at the ribs on the windward (lower) surface of the aeroshell.

Temperature contours[1] are also shown in Fig. 10-12 and display similar patterns to the Stanton number, with the higher temperatures at the nose and the ribs – which are exposed to higher heat fluxes – and lower temperatures on the panels. An exception to this is the *Shoulder* model, which is thinner at the shield edges, see Fig. 3a. Because of this, most of the thermal energy transferred at the rib's edges is conducted through the solid, leading to a local temperature increase, as shown in the temperature contours of the Shoulder model, see Fig. 10b and 10f. The technique used to obtain the experimental heat fluxes from the temperatures may not be accurate in predicting the heat fluxes at the edges of the Shoulder model, as the assumption of a semi-infinite body is violated. This implies that the increase in Stanton number predicted by the experiments at the shoulder of this geometry (see Fig. 7b) is non-physical and should be disregarded.

Figure 12 shows Stanton number and temperature profiles over the section passing along the ribs of each aeroshell at $\alpha = 0°$. While numerical results compare well with experiments, a consistent difference in the profiles is the Stanton number spike predicted by the CHT simulations at the edge of the aeroshells. These spikes are thought to be physical, as the supersonic expansion at the model shoulder shown in Fig. 6a leads to boundary layer thinning and therefore higher convective flux. In a previous CFD study of a faceted aeroshell at hypersonic speeds [6], similar heat flux spikes at the shoulder were shown, further confirming that the current experimental results do not capture the boundary layer thinning at the shoulder. In addition, the numerical heat flux over the sharp ribs is considerably higher than that predicted by the experiments, see Fig. 13a and 13b.

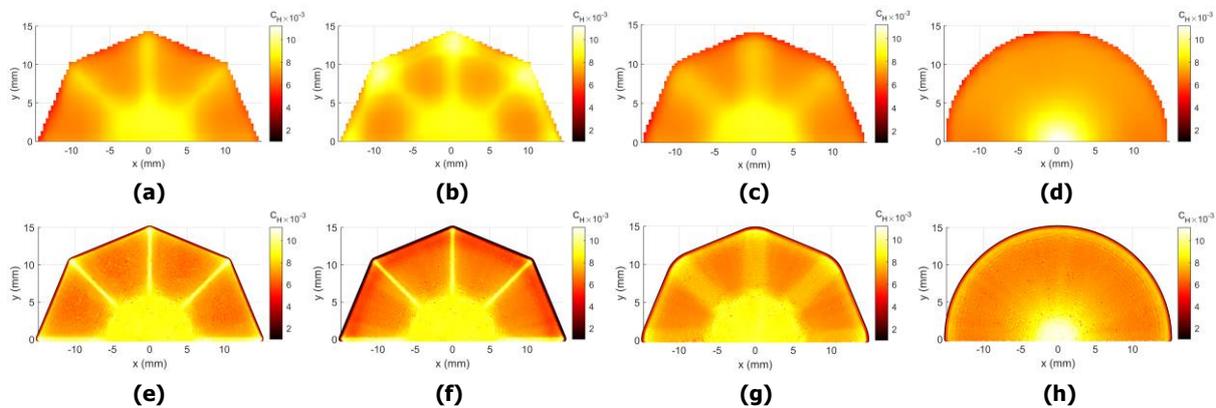

**Fig 7:** Experimental (above) and computational (below) Stanton number contours at $\alpha = 0°$, t = 5s. From the left: Sharp, Shoulder, Smooth, Sphere-cone

---

[1] To facilitate clear and meaningful comparisons between experimental and numerical temperature contours, experimental results are plotted at the time in each test which gives the closest match to those observed in simulations, which were always taken at t = 5s. This approach accounts for variations in initial temperature of the model and wind tunnel start-up transients between tests

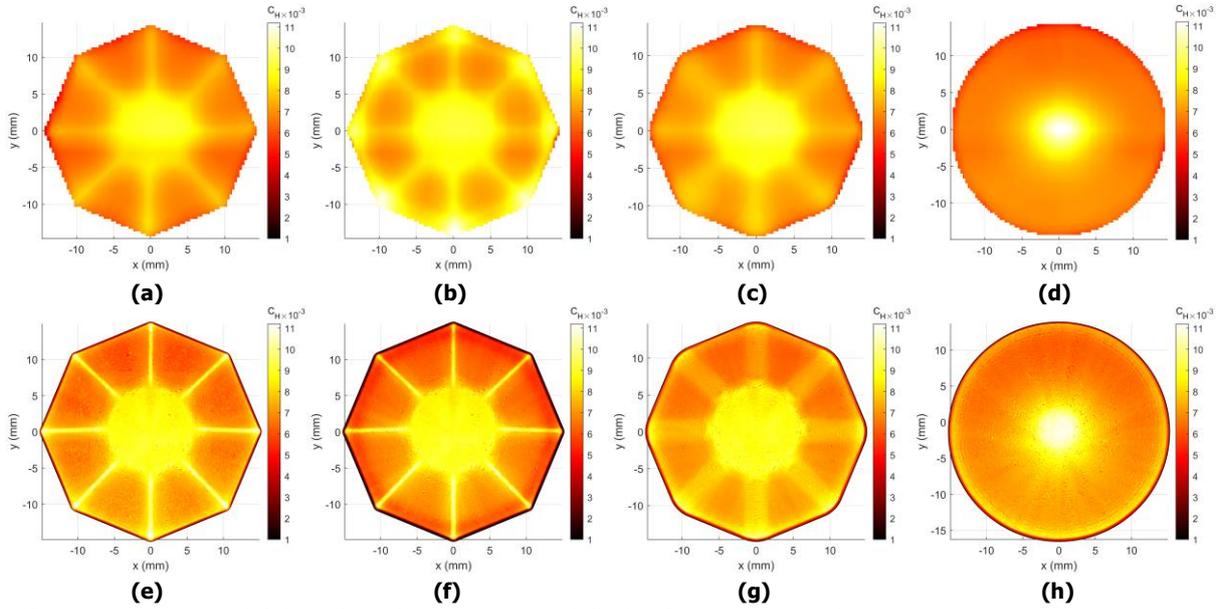

**Fig 8:** Experimental (above) and computational (below) Stanton number contours at α = 5°, t = 5s. From the left: Sharp, Shoulder, Smooth, Sphere-cone

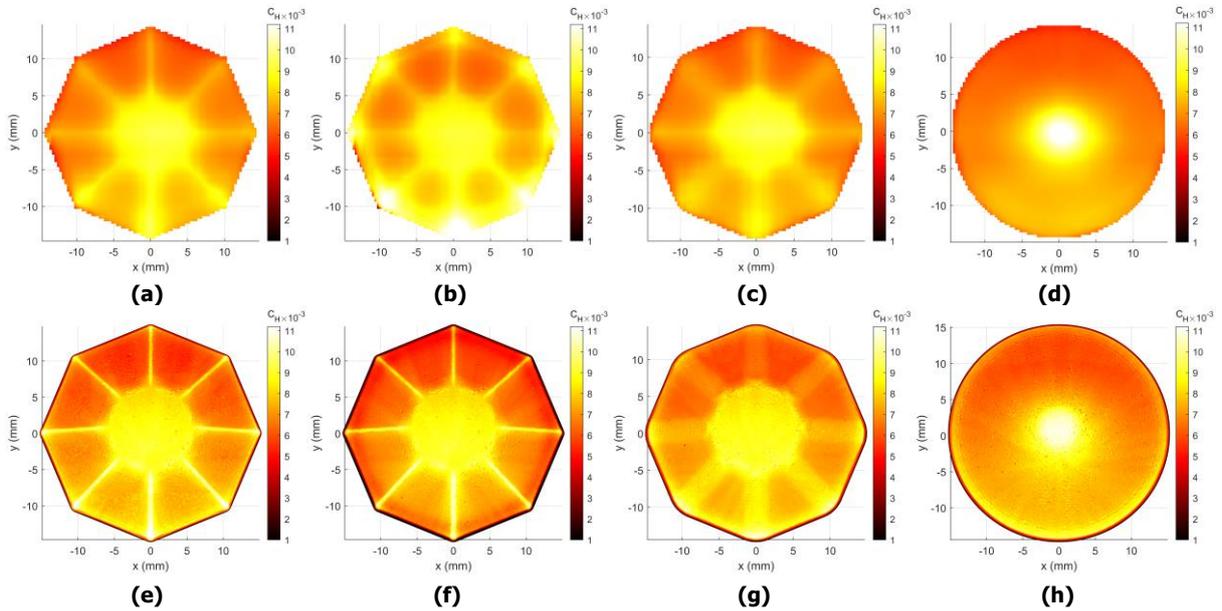

**Fig 9:** Experimental (above) and computational (below) Stanton number contours at α = 10°, t = 5s. From the left: Sharp, Shoulder, Smooth, Sphere-cone

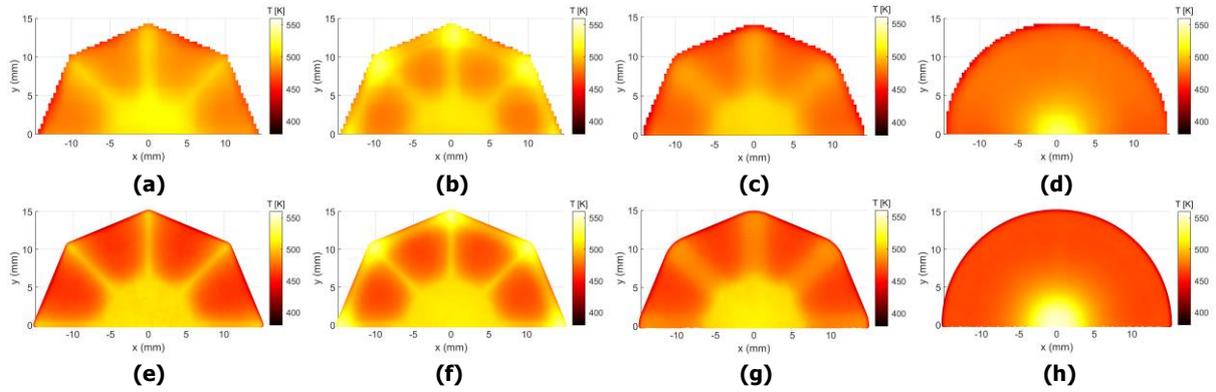

**Fig 10:** Experimental (above) and computational (below) temperature contours at α = 0°, t = 5s. From the left: Sharp, Shoulder, Smooth, Sphere-cone

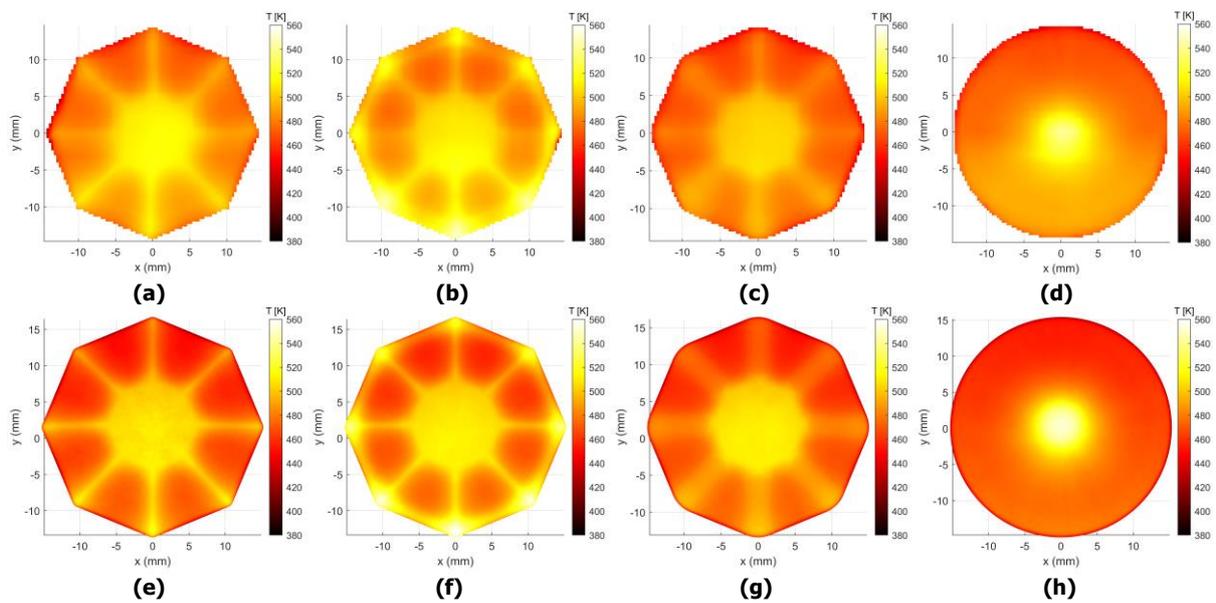

**Fig 11:** Experimental (above) and computational (below) temperature contours at α = 5°, t = 5s. From the left: Sharp, Shoulder, Smooth, Sphere-cone

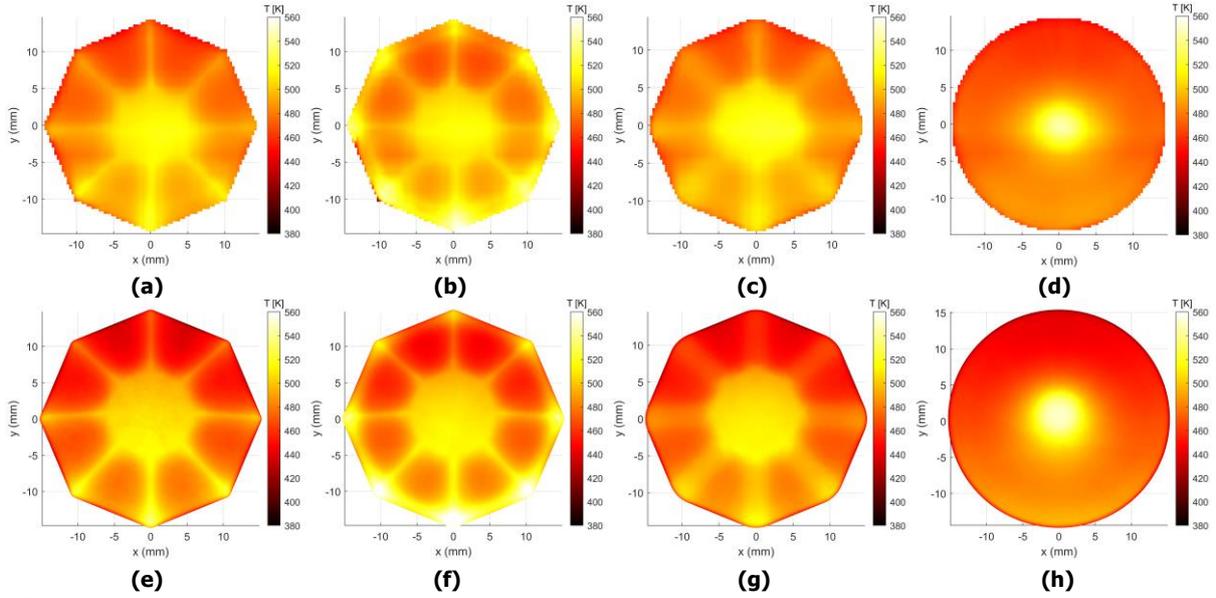

**Fig 12:** Experimental (above) and computational (below) temperature contours at α = 10°, t = 5s. From the left: Sharp, Shoulder, Smooth, Sphere-cone

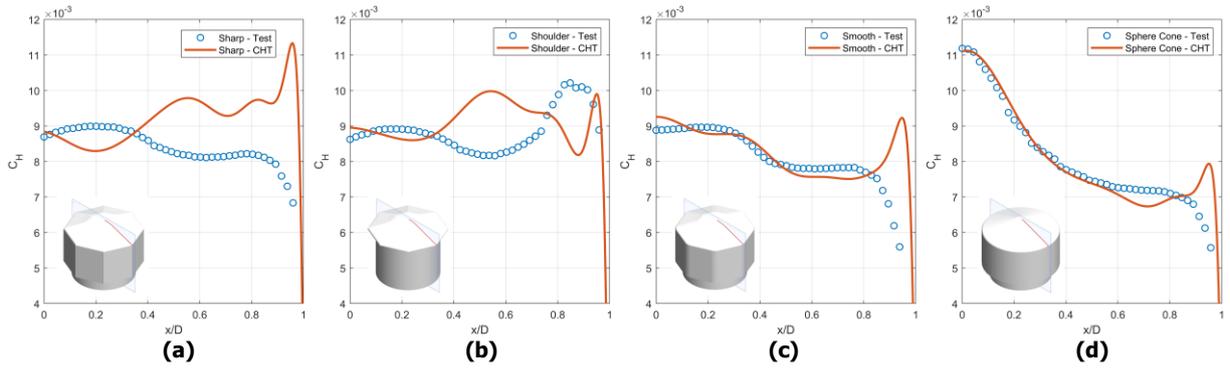

**Fig 13:** Stanton number profiles along the rib section at α = 0°, t = 5s. From the left: Sharp, Shoulder, Smooth, Sphere-cone

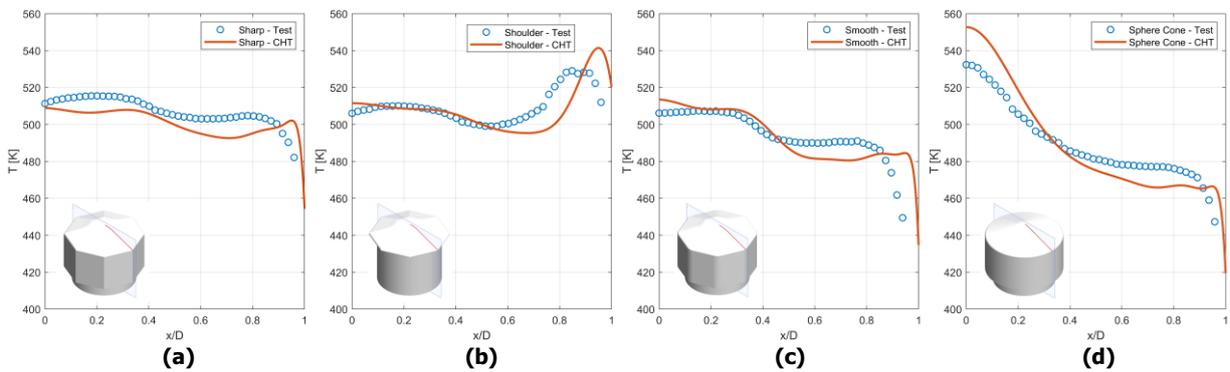

**Fig 14:** Temperature profiles along the rib section at α = 0°, t = 5s. From the left: Sharp, Shoulder, Smooth, Sphere-cone

### 3.3. Critical assessment of the results

The Mean Averaged Error, defined as:

$$\varepsilon_T = \frac{1}{n} \sum_{k=1}^{n} \frac{|T_{CHT}(k) - T_{exp}(k)|}{T_{exp}(k)} \qquad (2)$$

is evaluated at each experimental data point k shown in the profile plots. The Stanton number and temperature errors for each geometry and angle of attack are given in Table 1 and 2, respectively. Due to the CFD prediction of heat flux spikes at the aeroshell edges and higher numerical heat flux over sharp ribs, errors in Stanton number (8%) are, on average, higher than errors in temperature (2%). Experimental and numerical Stanton number for the *Smooth* and *Sphere-cone* geometries compare well, within 5%. In contrast, experimental and numerical results for geometries with sharp ribs do not compare well. In particular, Fig. 13 illustrates how experimental Stanton number values along the sharp (Fig. 13a and 13b) and smooth (Fig. 13c) ribs are very similar, while the numerical simulations predict up to 100% higher increase in Stanton number at the sharp ribs when compared to smooth ribs increase. There are different sources of error that can lead to this difference in results. Rees et al. [14] suggested that the highest global error in experimental Stanton number over faceted shapes at hypersonic speeds occurs at the sharp edges of the geometry, with values of up to 15% for heat fluxes obtained solving an Inverse Heat Conduction Problem (IHCP). In this study the heat fluxes were obtained solving the 1D heat conduction equation with a direct numerical technique, which is a less robust technique than IHCP and tends to magnify experimental errors and noise [13, 14]. In addition to the heat flux derivation, errors can arise from experimental sources, including optical calibration, uncertainty in MACOR emissivity at high temperatures (which occur at the sharp edges), as well as optical resolution. Rees et al. [15] numerically computed the Mach 5 flow over a cube and concluded that they were unsure whether or not the numerical Stanton number spikes at the sharp edges were physical, as numerical errors may arise when not modelling the material thermal response. However, the CHT analysis conducted in the present study effectively rules this out and thus suggests that the high heat flux spikes at the ribs are indeed a physical phenomenon which is just hard to capture experimentally.

In addition, experimental results cannot predict the heat flux spikes at the shoulder of each model, which are also thought to be physical, as described in Section 3.2. Solving the 1D heat equation with a direct numerical technique leads to considerable uncertainties, as transverse heat transfer within the geometry is neglected. As discussed in [14], heat conduction near corners and edges is highly three-dimensional and therefore needs to be modelled accordingly. It is likely that heat flux spikes at the model shoulders are thus not captured in the experiments due to the neglection of transverse heat transfer.

Table 1.    Stanton number error in %

| α [°] | Sharp | Shoulder | Smooth | Sphere Cone |
|---|---|---|---|---|
| 0 | 15.03 | 8.86 | 5.09 | 3.96 |
| 5 | 14.81 | 9.56 | 4.46 | 5.61 |
| 10 | 11.05 | 8.81 | 4.59 | 3.82 |

Table 2.    Temperature error in %

| α [°] | Sharp | Shoulder | Smooth | Sphere Cone |
|---|---|---|---|---|
| 0 | 1.45 | 1.15 | 1.28 | 1.91 |
| 5 | 1.02 | 1.46 | 1.18 | 2.12 |
| 10 | 0.89 | 1.63 | 1.75 | 2.45 |

## 4. Conclusions

Previous numerical studies of hypersonic flows over faceted aeroshells did not include the material thermal response of the solid [6, 7]. This can lead to errors in the computation of the heat fluxes over the geometry, as the wall boundary condition is not physical and thermal conduction through the solid is ignored. In this work a Conjugate Heat Transfer analysis of four different aeroshell models at Mach 5 is performed with the commercial solver STAR-CCM+, fully coupling the flow energy equation to the solid heat conduction equation. Results show that Stanton number increases considerably at the ribs between the facets, due to boundary layer thinning. The heat flux increase depends on the level of sharpness of the ribs, with smooth ribs leading to lower heat convection. Heat transfer predictions are compared with wind tunnel data and show good agreement. However, experimental results are not able to predict the heat flux spikes at the aeroshells shoulder, since the numerical technique used to obtain heat fluxes from experimental temperatures only solved the 1D heat conduction equation, while thermal conduction at the shoulders is highly three-dimensional. The largest discrepancies between CFD and experiments occur at the sharp ribs between the facets, with numerical heat fluxes being higher than experimental ones. These highly localised heat flux peaks at the ribs are thought to be physical and their absence in experiments is attributed to the limited spatial resolution of the IRT measurements, errors in optical calibration, and errors introduced in the heat flux derivation from temperature measurements.